\documentstyle [12pt] {article}

\begin{document}
\vspace*{0.5 cm}
\begin{center}
{\bf 
GETTING THE MOST FROM ATMOSPHERIC NEUTRINOS} \\
\vspace*{0.5 cm}
\vspace*{0.5 cm}
{\bf  J. Pantaleone}\\
\vspace*{0.5 cm}
Department of Physics and Astronomy\\
University of Alaska Anchorage\\
Anchorage, Alaska  99508 \\
\vspace*{0.5 cm}
{\bf ABSTRACT} \\
\end{center}

Observations of atmospheric neutrinos by the SuperKamiokande
collaboration have demonstrated large mixing
of the muon-neutrino.  
However the present atmospheric neutrino data does not
significantly constrain the associated mixing of the electron-neutrino, 
or the sign of the mass-squared difference.
Here we identify the diagnostics for these quantities
and they also test the theory of how matter affects neutrino oscillations.
These diagnostics are 
a dip in the sub-GeV muon flux at a zenith angle of 110 degrees,
a bump in the electron up-down asymmetry at multi-GeV energies and
a bump in the muon-antimuon upward asymmetry.

\newpage

Neutrinos produced in the atmosphere by cosmic rays have been observed by
several underground neutrino detectors
\cite{SK_obs}-\cite{at_other}.
These measurements have long suggested new neutrino physics \cite{1_osc}, 
and this has been conclusively demonstrated by recent data.
SuperKamiokande \cite{SK_obs} has observed a deficit of 
muon-neutrinos at large zenith angles
while the electron-neutrino ($\nu_e$) flux is as expected.  
These results are
well explained by large mixing between the muon-neutrino ($\nu_\mu$) 
and the tau-neutrino ($\nu_\tau$) \cite{SK_m}.

To interpret the atmospheric neutrino observations, 
we work in a framework using all three neutrino types \cite{foot}.
Three neutrino mixing depends on two mass-squared difference scales,
three mixing angles and a phase---however 
not all of these parameters are relevant for atmospheric neutrinos. 
The several solar neutrino observations \cite{solar}
suggest neutrino mixing also occurs on a scale 
below what can be resolved by the measurements of atmospheric neutrinos.
So for atmospheric neutrino analyses the smallest mass-squared difference
can be neglected, which then renders one mixing
angle and the phase unobservable.  The resulting framework has been used 
many times to find constraints on the neutrino mixing parameters
from the Kamiokande results \cite{jp_atmos}-\cite{3mat}
and also now the preliminary SuperKamiokande results \cite{SK1ms}.  
Here we adopt a different approach \cite{flp} than these latter references. 
We strive to qualitatively understand the present data, and to use this
understanding to determine which measurements are most important.

The primary motivation for this work is to find the best way to extract
the $\nu_e$ mixing parameter from the atmospheric neutrino data.
Binning the data as theory suggests 
can enhance the statistical significance of a signal,
and hence increase the sensitivity to $\nu_e$ mixing.
Also, a believable demonstration of nonzero $\nu_e$ mixing
will require a clear, unambiguous diagnostic.
The $\nu_e$ mixing is not a dominant feature in the current data, 
but it is an especially interesting parameter.
It will help determine what level of sensitivity
is needed by next generation double beta-decay experiments in order
to test if neutrinos are majorana or dirac fermions.  The $\nu_e$
mixing at the atmospheric scale is also necessary to  
interpret the solar neutrino measurements \cite{KP}.  
Constraints on $\nu_e$ mixing near the atmospheric
scale have recently been placed by the CHOOZ long-baseline 
reactor neutrino experiment \cite{CHOOZ}
and will be placed by the long-base line accelerator experiments 
(MINOS and K2K \cite{lbl}) which are presently under construction.
However these experiments may not be able to probe small enough mass scales.  
Even if they do observe some neutrino mixing effects,
they are relatively insensitive to
neutrino forward scattering off of the electron background 
\cite{MSW} of the Earth.
The sign of the neutrino mass-squared difference 
parameter can not be observed without both $\nu_e$ 
mixing and matter effects.  
We shall show that matter effects are {\it crucial} to
discerning the $\nu_e$ mixing from the atmospheric neutrino observations.
Thus these observations test how
the matter background affects neutrino mixing.
This is important because neutrino forward scattering 
effects have never been observed 
before, but they are heavily used in interpretations of solar neutrino 
observations.

The parameterization of the mixing 
\begin{equation}
| \nu_\alpha > \ \ = \ \ U_{\alpha i} | \nu_i >
\end{equation}
between the flavor eigenstates, \(\alpha = e, \mu, \tau\),
and the mass eigenstates, \(i = 1, 2, 3\), 
is here chosen to be
\begin{equation}
U = \left[ \begin{array}{ccc}
\cos \phi               &  0           & \sin \phi \\
- \sin \psi \sin \phi   & - \cos \psi  & \sin \psi \cos \phi \\
- \cos \psi \sin \phi   &   \sin \psi  & \cos \psi \cos \phi 
\end{array} \right]
\label{U}
\end{equation}
where the angles \(\phi\) and \(\psi\) range from 0 to $\pi/2$,
and the single mass-squared difference scale, $\Delta$, can be positive 
or negative.  This parameterization is the convention used in Ref. \cite{KP}, 
it was developed so that matter effects \cite{MSW} are straightforward. 
The oscillation amplitudes depend on the matter background. 
For arbitrary matter distributions, the amplitude
for oscillations from $\nu_\alpha$ to $\nu_\beta$ can be written as
\begin{equation}
{\rm Amp}_{\alpha \beta} = \left[ \begin{array}{ccc}
B_{e e}   & \sin \psi B_{x e}  & \cos \psi B_{x e} \\
\sin \psi B_{e x}  &   \cos^2 \psi  + \sin^2 \psi B_{x x}  
& \sin \psi \cos \psi ( B_{x x} - 1 ) \\
\cos \psi B_{e x} & \sin \psi \cos \psi ( B_{x x} - 1 ) & 
\cos^2 \psi B_{x x} + \sin^2 \psi  
\end{array} \right]
\label{B}
\end{equation}
Here the $B$'s are the conventional amplitudes for oscillations in the 
`standard' two-flavor approximation and satisfy the usual unitarity 
constraints:
$|B_{ee}|^2 = |B_{xx}|^2 = 1 - |B_{ex}|^2$.
They include all of the matter effects and the only mixing parameter they 
depends on is $\phi$.
Note that the $B$'s are different for neutrinos and antineutrinos.
We see that the three-neutrino, one mass scale approximation reduces to 
$\nu_e \leftrightarrow \nu_\tau$ oscillations in the limit $\sin \psi 
\rightarrow 0$, 
to $\nu_e \leftrightarrow \nu_\mu$ oscillations in the limit $\sin \psi 
\rightarrow 1$, and
to $\nu_\mu \leftrightarrow \nu_\tau$ oscillations in the limit $B_{ex} 
\rightarrow 0$.
Explicit expressions for the oscillation probabilities in this framework for
a constant density background can be found in Ref. \cite{jp_atmos}.

The detectors observe the electrons and muons produced by the atmospheric 
neutrinos.  
The ratio of the observed to expected charged lepton fluxes produced by 
neutrinos are given by
\begin{eqnarray}
\frac{N_e}{N_e^0} &=& 1 + \epsilon
\nonumber
\\
\frac{N_\mu}{N_\mu^0}& =& 1 - \frac{1}{2} \sin^2 2 \psi [ 1 - {\rm Re} 
( B_{x x} ) ]
			- \sin^2 \psi \frac{1}{r} \epsilon
\label{nratios}
\end{eqnarray}
where
\begin{equation}
\epsilon = ( r \sin^2 \psi  - 1 ) | B_{ex}|^2
\end{equation}
The expected ratio of the fluxes without oscillations is here denoted by 
$ r = N_\mu^0 / N_e^0 $.
$r$ varies somewhat with energy and zenith angle, $\theta$, but 
typically it is about 2 since the neutrinos are produced by 
$\pi^{\pm} \rightarrow \mu^{\pm} + \nu_\mu$ followed by $\mu^{\pm} 
\rightarrow e^{\pm} + \nu_\mu + \nu_e$.
The observation of a deficit of
atmospheric $\nu_\mu$ at large zenith angle while the $\nu_e$ flux is as 
expected
can both be explained by large `$\nu_\mu-\nu_\tau$' mixing, 
i.e. $\psi \approx \pi/4$.
This is because the oscillatory term ${\rm Re} ( B_{x x} )$ is one for short 
propagation distances
but averages to zero at large propagation distances,
while a cancellation produces a naturally small value for $\epsilon$. 
A crude, heuristic explanation of this cancellation in $\epsilon$ is possible.
Without mixing there are expected to be twice as many
$\nu_\mu$ as their are $\nu_e$'s, however when `$\nu_\mu-\nu_\tau$' mixing is 
maximal
only half of the $\nu_\mu$'s are left to mix with the $\nu_e$'s.  This leaves 
equal numbers of $\nu_\mu$'s and $\nu_e$'s so the `$\nu_e-\nu_\mu$' mixing is 
difficult to detect.   Thus the gross features of the SuperKamiokande 
atmospheric
neutrino data do not strongly constrain the $| B_{ex}|^2$ factor in 
$\epsilon$ or, 
consequently, the $\phi$ mixing parameter.

To diagnose the mixing with $\nu_e$, we consider independently 
three classes of atmospheric neutrino data: the sub-Gev contained events,
the multi-GeV events, and the `high' energy through going muons 
(which have an average energy of about 100 GeV).
The energy scale for the oscillations amplitudes is given by
\begin{eqnarray}
E_{mantle} &\approx& 6.7 {\rm GeV} \left( \frac{2.0 g/{\rm cm}^3}
{Y_e \rho}\right) 
\left( \frac{\Delta}{2 \times 10^{-3} {\rm eV}^2} \right)
\nonumber
\\
E_{core} &\approx& 0.4 E_{mantle} 
\label{e_res}
\end{eqnarray}
where \(Y_e\) is the number of electrons per nucleon,
and \(\rho\) is the density.
Note that neutrinos propagate through the core only at very large zenith 
angles, when $\cos ( \theta ) < -0.84$.
The behavior of the oscillation amplitude is qualitatively different 
in the three energy regimes.

{\it Sub-GeV neutrinos.}  Here $r \approx 2$ for all zenith angles \cite{flux}.
Taking $\psi \approx \pi/4$, as suggested by the observations, 
then the $\epsilon$ terms can be neglected because they are suppressed by
the cancellation.
The dominant effect of $\nu_e$ mixing is in the muon data!
For a constant density,  
\begin{eqnarray}
\frac{N_\mu}{N_\mu^0} \approx \frac{1}{2} [ 1 
&+& \cos^2 \phi \cos( 2 \pi \left( \frac{\Delta}{2 \times 10^{-3} 
{\rm eV}^2} \right)
\left( \frac{ 0.5 {\rm GeV} }{E}\right) \left( \frac{t}{620 {\rm km}}\right) )
\nonumber
\\
&+& \sin^2 \phi \cos ( 2 \pi \left( \frac{2.0 {\rm g}/{\rm cm}^3}{Y_e \rho}
\right) 
\left( \frac{t}{\rm 8300 {\rm km}} \right) \cos^2 \phi ) ]
\label{3mat}
\end{eqnarray}
for neutrinos {\it and} antineutrinos.   
This expression contains two oscillatory terms which are quite different.
The first is the `usual' two-flavor vacuum oscillations while the second
is a three-flavor effect which results because the matter background splits 
the degeneracy between the lightest neutrino states \cite{jp_atmos} \cite{3mat}.  
Detection of the matter induced oscillations would indicate mixing with the 
$\nu_e$.

The vacuum oscillations affect the angular distribution of sub-GeV neutrinos 
near the 
horizon, while `matter oscillations' affect neutrinos in the lower hemisphere.
To suppress the former, and to statistically enhance the resolution of the 
latter, 
it is natural to sum over all sub-GeV energies.
Fig. (1) shows the expected distortions in the zenith angle distribution of 
the sub-GeV muon data, with and without $\nu_e$ mixing.
The angular resolutions of the SuperKamiokande and SoudanII detectors have 
been used.
The most prominent effect of $\nu_e$ mixing is the dip in the flux 
where the matter oscillation term has its first minimum,
at $\cos \theta \approx (- 0.32 / \cos^2 \phi ) $.
The higher densities in the core act to increase the matter oscillation phase
and hence angular smearing suppresses the matter 
oscillations at the very largest zenith angles.
The angular resolution possible in a fine grained detector like SoudanII,
which can observe some of the nuclear recoil particles,
appears to be necessary to resolve the matter oscillations.
A future, larger, fine-grained atmospheric neutrino detector may be able 
to observe this signal.
The present sub-GeV data does not constrain the $\nu_e$ mixing parameter $\phi$.

Geomagnetic effects will complicate the resolution of this matter effect since 
they produce asymmetric angular variations in the neutrino flux.  
However the different effects can be disentangled,
in principle, since the matter oscillation is energy
independent, independent of azimuthal angle, and the same for
all detector sites, while the geomagnetic effect is not.
There is only one free parameter that describes the size and location of
the dip produced by matter oscillations, the $\nu_e$ mixing angle $\phi$.

{\it Multi-GeV neutrinos.}  
For nonzero $\nu_e$ mixing ($\phi \neq 0$), there is a two-flavor resonance 
\cite{MSW} with the
matter background at the energy in Eq. (\ref{e_res}).  To resolve this 
energy dependent effect,
it is natural to enhance statistics by summing over zenith angle to
form the up-down asymmetries \cite{SK_m}.  
The electron asymmetry may now show a signal because the cancellation in 
$\epsilon$ may no longer be complete
since at multi-GeV energies $r$ is considerably larger than 2 for $\cos 
\theta \approx \pm 1 $,  
and because $|B_{e x}|^2 \approx 0.5$ for neutrinos (or antineutrinos) with 
energies near $E_{mantle}$.   
The curves in Fig. (2) shows the electron up-down asymmetry, 
$A_e = (U-D)/(U+D)$, as a function of energy,
for the angular resolution of SuperKamiokande.
The $\nu_e$ mixing generally causes a positive bump at the resonance energy.
However the size of this bump is quite sensitive to the precise value of $\psi$.
A chi-squared fit to this SuperKamiokande data 
shows that the multi-GeV data does not significantly constrain the the 
$\nu_e$ mixing parameter $\phi$
for the values of $\Delta$ and $\psi$ allowed by the SuperKamiokande data. 
However some constraints may be possible with higher statistics,
and a close examination of systematic errors (such as particle 
misidentification) 
at the high energy end of the multi-GeV electron up-down asymmetry data.

The effects of $\nu_e$ mixing and matter are more difficult 
to observe in the `usual' multi-GeV muon up-down asymmetry.
This is because the muon asymmetry already has large distortions from the
large `$\nu_\mu-\nu_\tau$' vacuum oscillations, and because 
the oscillatory terms in Eq. (\ref{3mat}) tend to be suppressed near 
resonance energies by energy and angular smearing.
However if multi-GeV muon's and antimuon's can be discriminated in 
SuperKamiokande \cite{LoSecco} or a future
atmospheric neutrino detector, this would provide an especially interesting, 
additional diagnostic.

The asymmetry between muons and antimuons in the upward fluxes 
$A_{\mu-{\bar \mu}} = ((U_\mu/D_\mu) - (U_{\bar \mu}/D_{\bar \mu})) 
/ ((U_\mu/D_\mu) + (U_{\bar \mu}/D_{\bar \mu}) )$, 
would show a bump at the resonance energy.
The size of the bump in $A_{\mu-{\bar \mu}}$ is comparable to the size of 
the bump in $A_e$.
The bump is produced by $\nu_e$ mixing and matter effects. 
$A_{\mu-{\bar \mu}}$ is particularly interesting 
because it is especially sensitive to the sign of the neutrino mass-squared 
parameter, $\Delta$.  
Flipping the sign of $\Delta$ does not affect the sub-GeV diagnostic at all,
it reduces the size of the bump in the multi-GeV electron
up-down asymmetry by 40\%, but it flips the sign of the bump in the 
muon-antimuon asymmetry.
$\Delta$ is positive if neutrinos have a mass 
heirarchy similar
to the charged leptons, but this must be experimentally determined.

{\it `High' energy neutrinos.}
For $E_{mantle} < < E$, matter effect suppress all trace of mixing with 
the $\nu_e$.
The leading term in the two flavor amplitude vanishes as
\begin{equation}
| B_{ex} |^2 \rightarrow \frac{1}{2} \left( \frac{E_{mantle}}{E} \right)^2 
\sin^2 2 \phi
\end{equation}
for neutrinos and antineutrinos.
In this energy regime, the three-flavor framework reduces exactly to
the `usual' two-flavor `$\nu_\mu - \nu_\tau$' vacuum oscillation expression.
Thus the through going muon data at SuperKamiokande, MACRO, AMANDA, etc. 
is completely insensitive to the mixing with $\nu_e$.

\begin{center}
Acknowledgements
\end{center}

I would like to thank T.K. Kuo, J. Learned and S. Pakvasa 
for useful discussions.  
This work is supported in part by Research Corporation.

\raggedbottom

\raggedbottom

\newpage

\newpage

\begin{center}
{\bf FIGURE CAPTION} \\
\end{center}
\vspace*{0.6cm}

\noindent {\bf Fig. 1.} The sub-GeV muon flux  ($N_\mu / N_\mu^0$) 
as a function of zenith angle.  
Points with error bars are current experimental measurements and
curves show expected distributions, 
where circle (triangle) markers denote SuperKamiokande \cite{SK_obs} 
(Soudan II \cite{at_other}) quantities.
The mixing parameters for the curves are
compatible with limits from CHOOZ and SuperKamiokande,
$\Delta = 1 \times 10^{-3}$, $\psi = \pi/4$ and $\phi = 0$ (lines with 
markers),  and $\phi = 0.50$ (lines without markers).

\noindent {\bf Fig. 2.}  Up-down electron asymmetry as a function of energy.
Points with error bars are current SuperKamiokande \cite{SK_m} measurements,
the thick (thin) curve shows the expected distribution
for $\Delta = 2 \times 10^{-3}$, $\psi = \pi/4$ ($\psi = \pi/5.5$) and 
$\phi=0.25$ (compatible with limits from CHOOZ and SuperKamiokande).
The SuperKamiokande angular resolution has been assumed.

\raggedbottom


\newpage

\begin{thebibliography}{99}

\bibitem{SK_obs}
Super-Kamiokande Coll., Y. Fukuda et al., hep-ex/9803003, 
hep-ex/9805006.

\bibitem{at_other}
Kamiokande Coll., K.S. Hirata et al., Phys. Lett. B {\bf 205}, 416 (1988);
Phys. Lett. B {\bf 280}, 146 (1992);
IMB Coll., R. Becker-Szendy et al., Phys. Rev. D {\bf 46}, 3720 (1992);
D. Casper et al., Phys. Rev. Lett. {\bf 66}, 2561 (1991);
Soudan II Coll., W.W.M. Allison et. Al., Phys. Lett. B {\bf 391}, 491 (1997).


\bibitem{1_osc}
J.G. Learned, S. Pakvasa and T.J. Weiler, Phys. Lett. B {\bf 207}, 79 (1988);
V. Barger and K. Whisnant, Phys. Lett. B {\bf 209}, 365 (1988);
K. Hidaka et al., Phys. Rev. Lett. {\bf 61}, 1537 (1988).

\bibitem{SK_m}
Super-Kamiokande Coll., Y. Fukuda et al., hep-ex/9807003.

\bibitem{foot}
This framework is incompatible with the observations of neutrino mixing by
the LSND short baseline experiment,
C. Athanassopoulos {\it et al.}, Phys. Rev. Lett. {\bf 75}, 2650 (1995).
However the experimental evidence from LSND is weak,
especially in light of the preliminary results from KARMEN, talk by
B. Zeitnitz at Neutrino-98, Takayama, Japan, June 1998.


\bibitem{solar}
Super-Kamiokande Coll., talk by Y. Suzuki at Neutrino-98, Takayama, Japan, 
June 1998;
B.T. Cleveland {\it et al.}, Nucl. Phys. B (Proc. Suppl.) {\bf 38}, 47 (1995);
Kamiokande collaboration, Y. Fukuda {\it et al.}, Phys. Rev. Lett., {\bf 77}, 
1683 (1996);
GALLEX Coll., W. Hampel {\it et al.}, Phys. Lett. {\bf B388}, 384 (1996);
SAGE Coll., J.N. Abdurashitov {\it et al.}, Phys. Rev. Lett. {\bf 77}, 
4708 (1996).


\bibitem{jp_atmos}
J. Pantaleone, Phys. Rev. D {\bf 49}, 2152 (1994);
J. Pantaleone, Phys. Lett. B {\bf 292}, 201 (1992).
N.B. the mixing convention in the current paper is identical to these 
references,
except that the labels on the 1 and 2 mass eigenstate are interchanged.

\bibitem{K1ms}
S. Midorikawa, M. Honda, K. Kasahara, Phys. Rev. D {\bf 44}, 3379 (1991);
C.Y. Cardall and G.M. Fuller, Phys. Rev. D {\bf 53}, 4421 (1996);
M. Narayan, G. Rajasekaran, S.Uma Sankar, Phys.Rev. D {\bf 56}, 437 (1997).
 
\bibitem{3mat}
G.L. Fogli, E. Lisi, D. Montanino, and G. Scioscia, Phys. Rev. D {\bf 55}, 
4385 (1997);
C. Giunti, C.W. Kim, M. Monteno, Nucl.Phys. B {\bf 521}, 3, (1998).


\bibitem{SK1ms}
O. Yasuda, hep-ph/9804400;
T. Teshima and T. Sakai, hep-ph/9805386;
V. Barger, T.J. Weiler, and K. Whisnant, hep-ph/9807319.

\bibitem{flp}
J.W. Flanagan, J.G. Learned, and S. Pakvasa,
Phys. Rev. D {\bf 57}, 2649 (1998). 


\bibitem{KP}
T.K. Kuo and J. Pantaleone, Rev. Mod. Phys. {\bf 61}, 937 (1989).


\bibitem{CHOOZ}
CHOOZ Coll. M. Apollonio {\it et al.},  Phys. Lett. B {\bf 420}, 397 (1998).



\bibitem{lbl}
MINOS Coll., "Neutrino Oscillation Physics at Fermilab: The NuMI-MINOS 
Project", NuMI-L-375, May (1998);
K2K Coll., K. Nishikawa, talk at Neutrino-98, Takayama, Japan, June 1998.



\bibitem{MSW}
L. Wolfenstein, Phys. Rev. D {\bf 17}, 2369 (1978); D {\bf 20}, 2634 (1979);
S. P. Mikheyev and A. Yu Smirnov, Yad. Fiz. {\bf 42}, 1441 (1985)
[Sov. J. Nucl. Phys. {\bf 42}, 913 (1985)]; Nuovo Cim. C {\bf 9}, 17
(1986).



\bibitem{flux}
M. Honda, T. Kajita, K. Kasahara and S. Midorikawa, Phys. Rev. D {\bf 52} 
4985 (1995);
V. Agrawal, T.K. Gaisser, P. Lipari and T. Stanev, Phys. Rev. D {\bf 53}, 
1314 (1996).

\bibitem{LoSecco}
J.M. LoSecco, hep-ph/9806318.



\end{thebibliography}
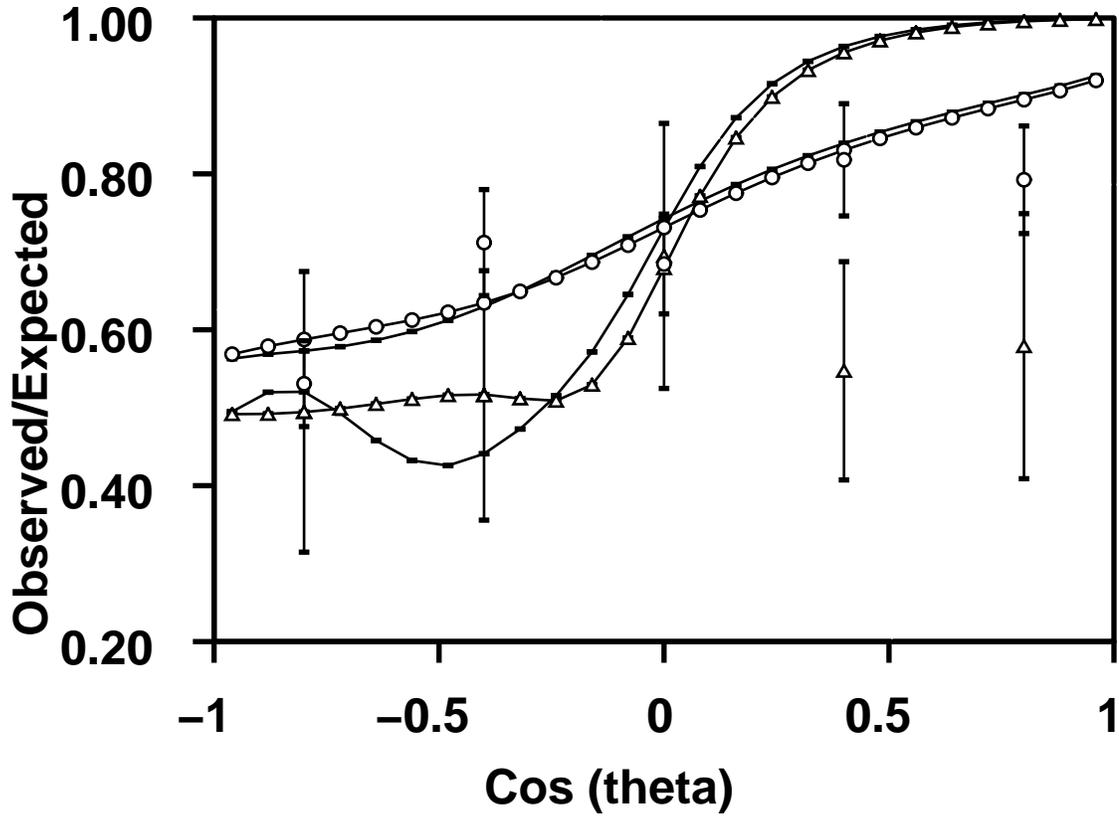
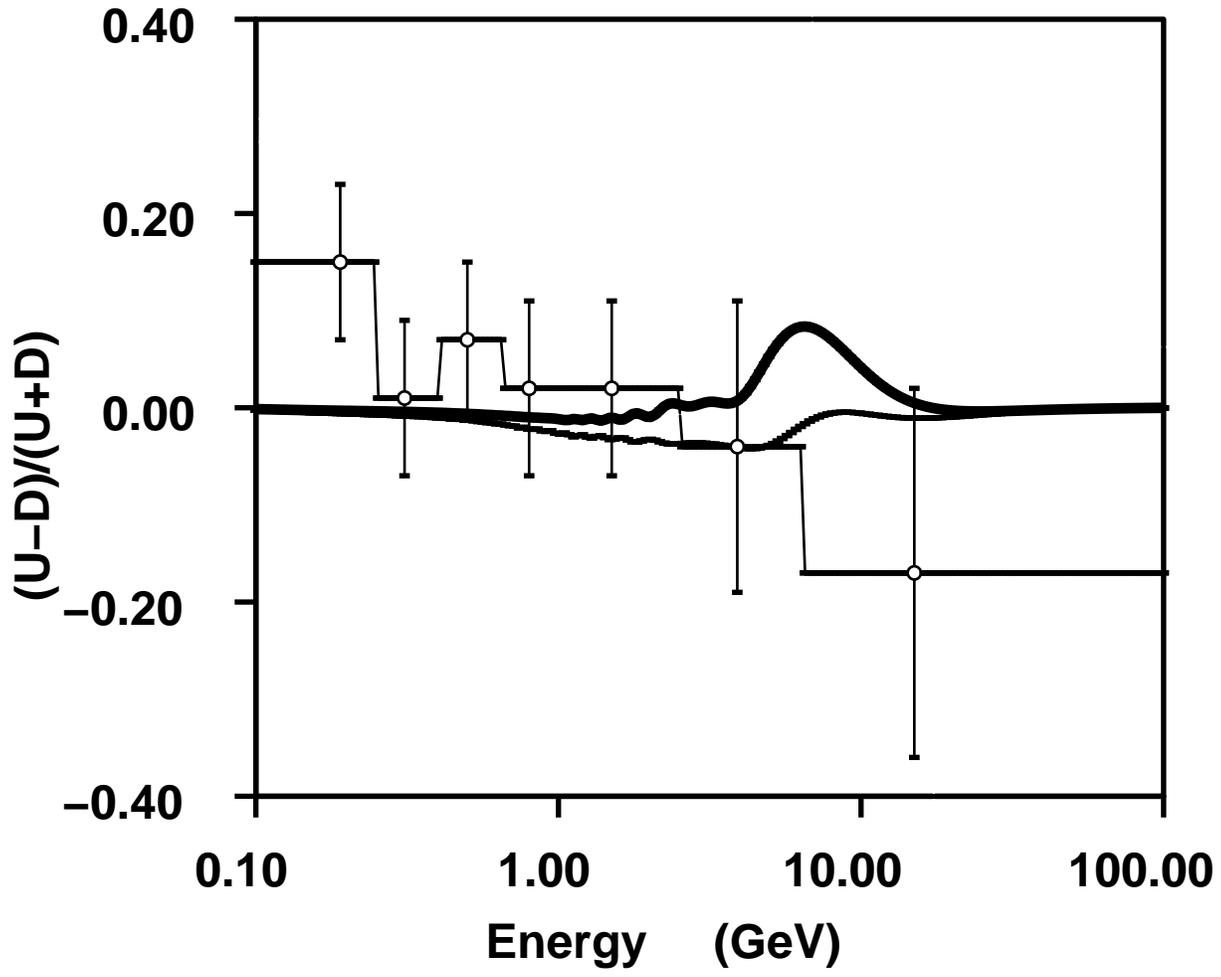
\end{document}